\documentclass{elsart}
\journal{NIM}

\usepackage{epsfig}

\usepackage{amssymb}
\usepackage{amsmath,amsfonts,verbatim,graphicx,float}
\usepackage{subfigure}
\usepackage{cite}
\pdfoutput=1

\newfont{\tensy}{cmsy10}

\begin{document}
 
\begin{frontmatter}



\title{Implementation of the Random Forest Method for the
Imaging Atmospheric Cherenkov Telescope MAGIC}

 \author[a]{J.~Albert}, 
 \author[b]{E.~Aliu}, 
 \author[c]{H.~Anderhub}, 
 \author[d]{P.~Antoranz}, 
 \author[b]{A.~Armada}, 
 \author[d]{M.~Asensio}, 
 \author[e]{C.~Baixeras}, 
 \author[d]{J.~A.~Barrio},
 \author[f]{H.~Bartko}, 
 \author[g]{D.~Bastieri}, 
% ...

 \address[a]{Universit\"at W\"urzburg, D-97074 W\"urzburg, Germany}
 \address[b]{Institut de F\'\i sica d'Altes Energies, Edifici Cn., E-08193 Bellaterra (Barcelona), Spain}
 \address[c]{ETH Zurich, CH-8093 Switzerland}
 \address[d]{Universidad Complutense, E-28040 Madrid, Spain}
 \address[e]{Universitat Aut\`onoma de Barcelona, E-08193 Bellaterra, Spain}
 \address[f]{Max-Planck-Institut f\"ur Physik, D-80805 M\"unchen, Germany}
 \address[g]{Universit\`a di Padova and INFN, I-35131 Padova, Italy} 
% ...

 \corauth[cor1]{Corresponding author.}

\begin{abstract}
The paper describes an application of the tree classification method Random Forest (RF),
as used in the analysis of data from the ground-based gamma telescope MAGIC. 
In such telescopes, cosmic $\gamma$-rays are observed and have to be discriminated
against a dominating background of hadronic cosmic-ray particles. 
We describe the application of RF for this gamma/hadron separation. 
The RF method often shows superior performance in comparison with 
traditional semi-empirical techniques.
Critical issues of the method and its implementation are discussed.
An application of the RF method for estimation of a continuous parameter
from related variables, rather than discrete classes, is also discussed.
\end{abstract}


\end{frontmatter}

\section{Introduction}
Ground-based gamma-ray astronomy has in recent years shown to be a 
source of  spectacular discoveries, 
constraining the evolution of the universe and contributing 
to the understanding of the origin of cosmic rays. 
Observations are based on Imaging Atmospheric Cherenkov Telescopes 
(IACTs), which take advantage of
the Cherenkov radiation emanating from the electromagnetic showers 
that develop during the absorption of gamma-rays 
in the atmosphere. The faint Cherenkov light flashes are collected in 
a large-diameter mirror, and recorded in a pixelized camera. 

Several IACT systems are in successful
operation today, both in the Northern (MAGIC, VERITAS) and Southern 
(HESS, CANGAROO) hemisphere; all but MAGIC are implemented as multi-telescope arrays. 
Their scientfic goals include galactic and extragalactic sources:
Supernova remnants, Pulsars, X-ray binaries, Microquasars,
Active Galactic Nuclei (blazars or radio galaxies), Starburst galaxies
and potentially also Gamma Ray Bursts. Due to their small aperture IACTs can only 
perform scans over small areeas, and usually concentrate 
on sources that have been identified at other wavelengths; however, the number of 
known gamma-ray emitters is increasing fast, and they provide essential
contributions to the understanding of the non-thermal universe. 

Events seen by an IACT have a very short ($\approx 2ns$) duration, and 
the shower image is recorded as a compact cluster of pixels
in the camera of the IACT. A principal component analysis
permits to express the characteristics of this cluster in image parameters,
which will present statistically different properties for the
(interesting) gamma-rays and the (dominating) hadronic background.
IACTs provide raw data with a signal-to-noise ratio much smaller than  $1\%$, 
even for bright gamma sources. Establishing powerful methods of hadronic 
background rejection thus is a prerequisite for the effective utilization of 
observations with the Cherenkov technique. The fact was recognized with the advent of
the IACT technique, and has been given ample room in the literature, both for telescope
arrays and single telescopes, e.g. \cite{hillas,hillas1,fegan,aharonian,kraw}. 
Multivariate methods using global test statistics 
(e.g. likelihood ratios or artificial neural networks) are specifically mentioned in 
\cite{fegan} and \cite{kraw}.

A case study for and comparison of different advanced classification methods 
for a single-dish IACT can be found in \cite{bock}. In the same article the main 
features of Cherenkov images measured by gamma-ray telescopes are addressed 
and explained, and the image parameters used in the $\gamma$/h separation are defined.

In this paper, largely derived from chapter 5 of \cite{hengst}, we limit ourselves to the 
implementation, usage, and functionality of the RF method for the single-dish system
MAGIC \cite{lorenz}. In \cite{hengst}, a more detailed discussion of the RF method
and comprehensive MC studies are given.
The implementation closely follows the method desribed
by L. Breiman \cite{breiman1}. The application 
in $\gamma$/h separation is discussed in detail. Recent MAGIC publications
(e.g. \cite{albert1, albert2, albert3} use the RF technique, and \cite{albert4}
dicusses it in the context of the reference observations of the Crab nebula.
A short comparative study with the 
established method of cuts in scaled image parameters is given. 
We also discuss an application of the RF method in estimating the 
gamma energy, a continuous 
variable, in terms of the observed image parameters. 
In the following chapter \ref{sec_basic} the Random Forest 
method will be described in 
detail, since existing mathematical treatments show only few practically 
useful aspects, if any. 
The reader not interested in these details may regard RF as a 
black-box tree classification method, and continue 
with the results in section \ref{results}. 

\section{Basics of the Random Forest (RF) method}
\label{sec_basic}
The Random Forest method is based on a collection of decision trees, built
up with some elements of random choices.
Like many other classification and regression methods, a Random Forest  
is constructed on the basis of training samples suitable for the application.
For the purpose of $\gamma$/h separation, the training samples contain the two 
classes of gammas (usually Monte Carlo (MC) data) 
and hadrons (usually OFF data, also ON\footnote{ON and OFF data are telescope 
data obtained by pointing at the source or on a nearby, sourceless region of the sky, 
respectively} or MC data are possible).
In the further discussion, the following definitions will be used:
We call the elements of the training sample {\it events}. 
Each event is characterized by a vector whose components are {\it image parameters}
obtained by analyzing the camera pixels. We use the familiar Hillas parameters \cite{hillas} 
and some additional parameters, but also
observation- and detector-related parameters, like $cos(\theta)$, 
$\theta$ being the zenith angle of the source. 
The space spanned by the event vectors is multi-dimensional. One can consider the 
training samples of gammas and hadrons
as a single labeled training sample, viz. each event has an integer label 
(called {\it hadronness}) indicating if the event belongs to the class of gammas (hadronness 0) 
or to the class of hadrons (hadronness 1).

\begin{figure}[h]
\begin{minipage} [c] {0.50\textwidth}
\includegraphics[totalheight=4.0cm]{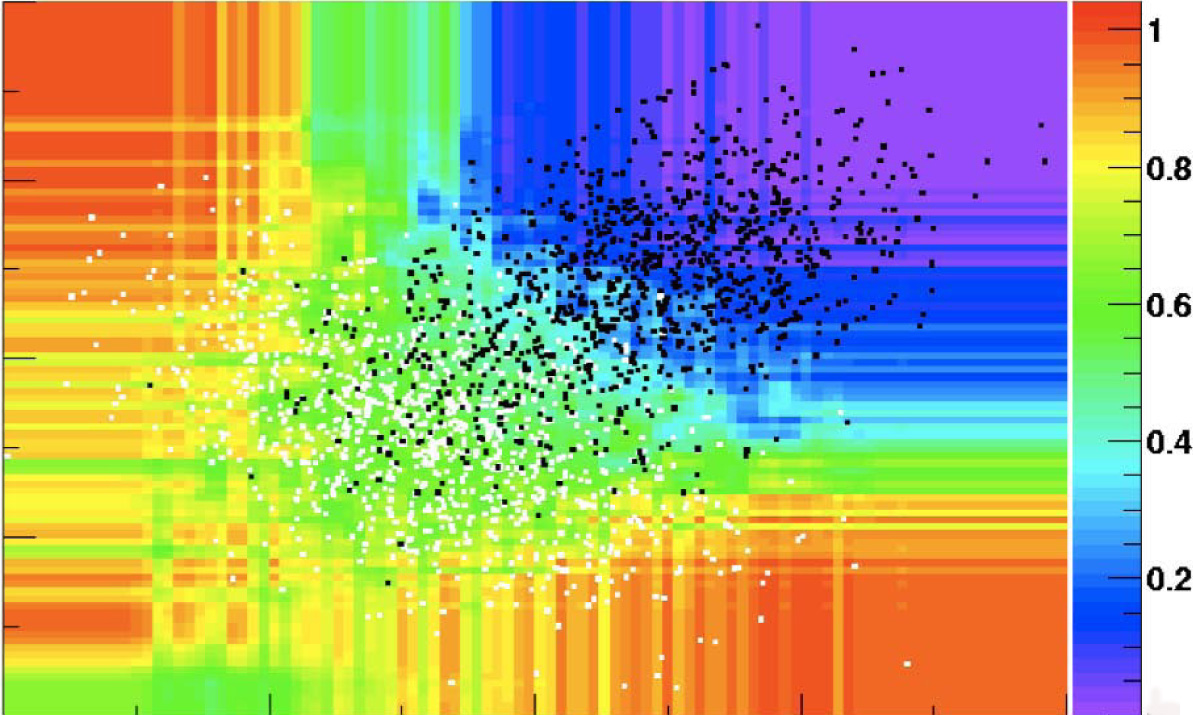}
\end{minipage}
\begin{minipage} [c] {0.5\textwidth}
\includegraphics[totalheight=4.5cm]{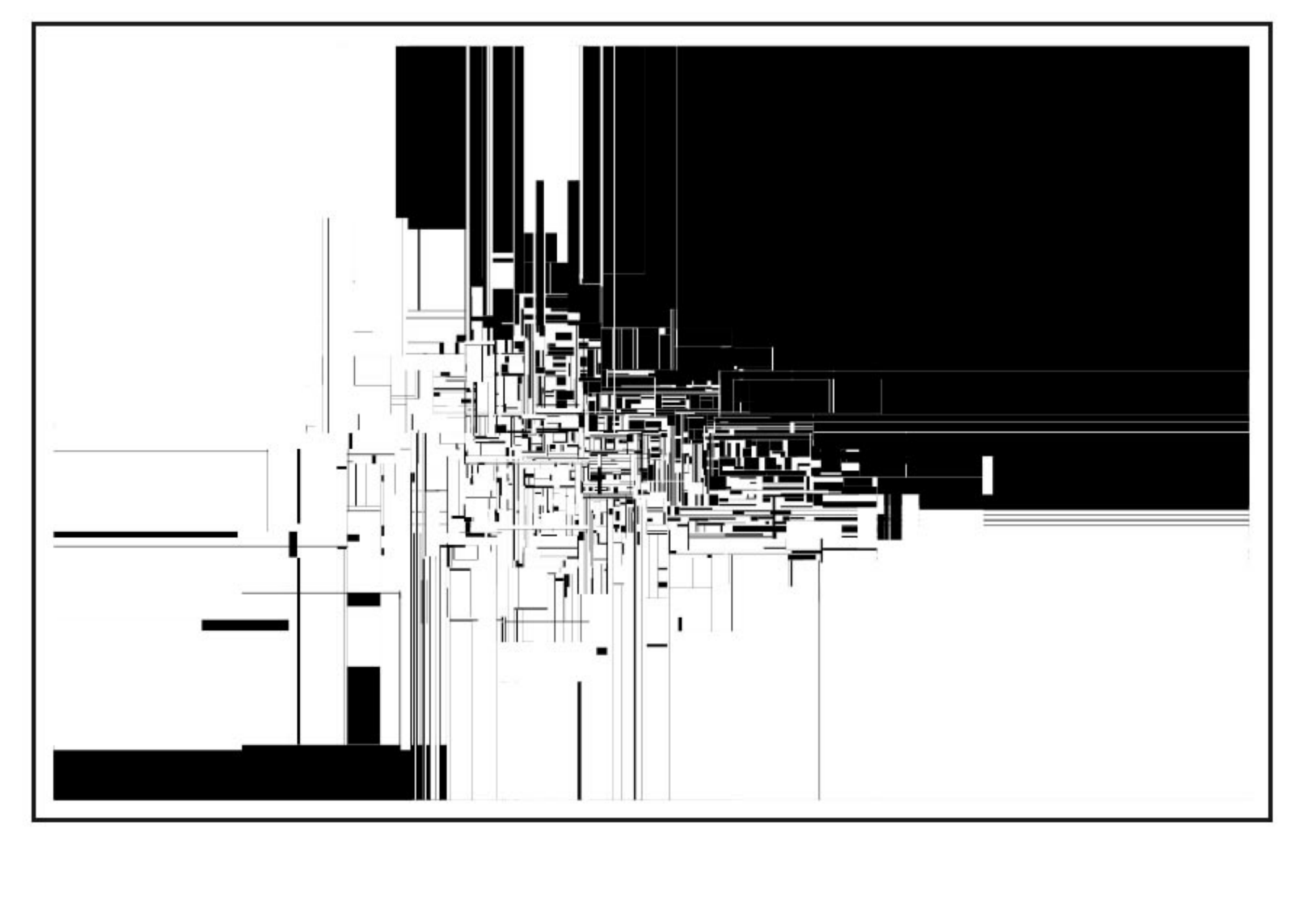}
\end{minipage}
\caption{{\it Left: Illustration of the RF method for a simple 2-dimensional model case. 
The black and white points are the observed points in class gamma and hadrons, 
respectively. They are distributed according to two different, but overlapping 2-dimensional
Gaussians. The result of separation in terms of hadronness is shown in colour. 
Right: The result of using a single tree on the same data gives no probability 
measure like hadronness, but only y/n answers. Its performance is inadequate.}}
\label{fig_pattern1}
\end{figure}

From this sample, a binary decision tree can be constructed, subdividing the parameter 
space first in two parts depending on one of the parameters, and subsequently repeating 
the process again and again for each part. The best choice of parameter and the
criteria for subdividing are discussed below.
Using a single tree for classification purposes, however, usually gives mediocre results. 
The tree is overoptimized on the training sample, and there is only poor generalization 
viz. new events will be classified rather badly. 
This is shown in figure~\ref{fig_pattern1}. Note, however, that even a set of trees 
(forest) results in some sparsely populated areas, where the hadronness 
necessarily is
not well defined, and the probability of misclassification may be substantial.

There is no pruning (tree simplification by removing some branches considered irrelevant)
of the trees in the Random Forest algorithm. Instead, the RF creates
a set of largely uncorrelated trees, and combines their results to form a 
generalized predictor. Two random elements prior to and within the tree growing process serve to 
approximate ideally uncorrelated trees; they are described in the following sections.

\subsection{Bootstrap aggregating (bagging)}
\label{subsec_bagging}
There is usually a single data sample in each class used for training. 
A straightforward solution to obtain independent trees is to 
split the training sample into as many non-overlapping subsamples as trees should be grown. 
However, there are usually not enough training data available for this approach. This is especially the case if dealing with 
air shower data, which are always costly to generate (w.r.t. computer time and storage space).
A different way is to produce a bootstrap sample for each tree by sampling n times with replacement from the 
original training sample containing n events. This procedure guarantees that the events' image parameter 
distributions are statistically identical for all bootstrap samples 
(and equal to the image parameter distributions of the 
original training sample, since the probability of selecting an event is constantly 1/n for the sampling with replacement 
procedure), while the bootstrap samples do not contain the same events. It may (and will) happen that certain 
events are taken more than just once:
The probability of not selecting a certain event is equal to $(1 - 1/n)$, 
which becomes $(1-1/n)^n$ when repeating the 
selection process n times. As $lim_{n\rightarrow\infty} (1 + x/n) = e^x$, 
the probability of not selecting an
event in the bootstrap procedure becomes $e^{-1}\approx1/3$. Thus, in each bootstrap sample there will be on average 
$(1 - 1/e)$  original training events, the rest (also kept in the sample) are copies.

\subsection{Tree growing and random split selection}
The tree growing begins with the complete sample contained in a single node, the so called root node, 
which is identical to the complete image parameter space. In the following the $\gamma$/h separation is achieved by 
splitting (or cutting) each node into two successor nodes using one of the image parameters at a time, with a 
cut value optimized to separate the sample into its classes (in our case two: gammas and hadrons). This corresponds to a 
successive division of the image parameter space into hypercubes.
In order to measure the classification power (separation ability) of an image parameter and to 
optimize the cut value, the Gini index is used The Gini index is a frequently used 
measure in dealing with classifiers, originally in economics. Named after the Italian economist 
Corrado Gini,
it measures the inequality of two distributions, 
e.g. gamma acceptance and hadron acceptance as function of a cut in a variable. 
It is defined as the ratio between a) the area spanned
by the observed cumulative distribution and the hypothetical cumulative distribution 
for a non-discriminating variable (uniform distribution, 45-degree line), and b) the 
area under this uniform distribution. It is a variable between zero and one;
a low Gini coefficient indicates more equal distributions, a 
high Gini coefficient shows unequal distribution. 
 
The choice of the parameter 
taken for splitting is randomized (see below for details).
The splitting process stops if the node size (events per node) falls below a limit specified by the 
user, or if there are only events of one class (only gammas or only hadrons) left in the node, which 
therefore needs not be split further. 
These terminal nodes can also be called elementary hypercubes, they cover the entire image parameter space 
without intersections or gaps. To each terminal node the remaining training events assign a 
class label $l$ (0 for gammas, 1 for hadrons).
For terminal nodes still containing a mixture of events of different classes, 
a mean value is calculated for $l$, taking into account the 
class populations $N_h$ of hadrons and $N_{\gamma}$ of gammas: $l = N_h / (N_h + N_{\gamma})$.
The original program \cite{breiman2} uses a majority vote, and does not calculate mean values.

\begin{figure}[h]
\begin{center}
\includegraphics[totalheight=5cm]{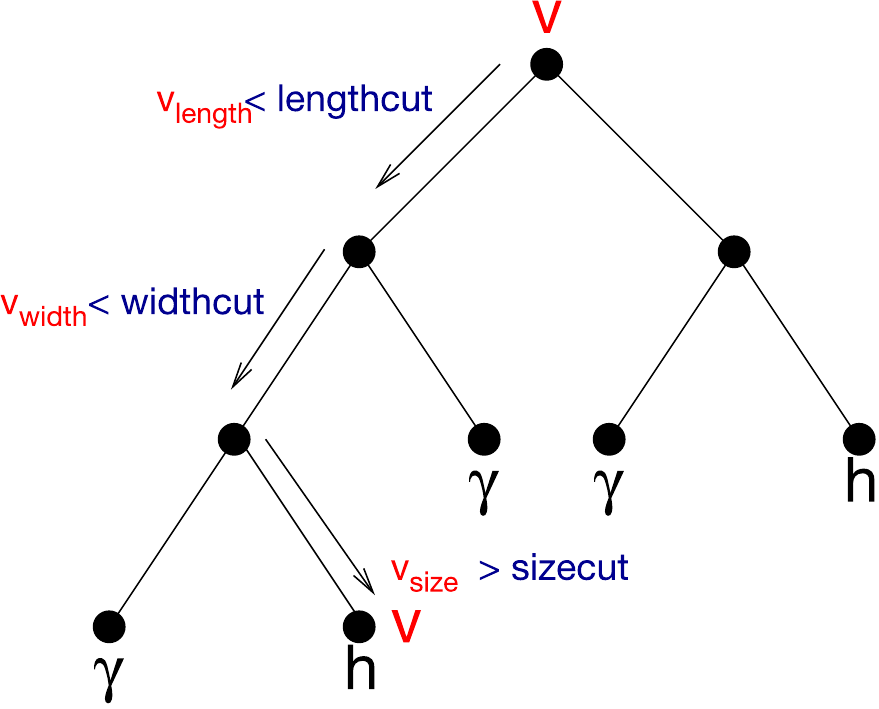}
\caption{{\it Sketch of a tree structure for the classification of an event $v$ with
components $v_{length}$, $v_{width}$, and $v_{size}$. 
One can follow the decision path through the tree, 
leading to classification of the event as hadron.}}
\label{fig_tree}
\end{center}
\end{figure}

Before going into more details, the classification process is briefly described:
One can take a completely grown tree as starting point 
(see figure~\ref{fig_tree}). 
The task is to classify an event
characterized by a vector $v$ in the image parameter space. $v$ is fed into the decision tree; 
at the first (highest level) node 
there is a split in a certain image parameter (e.g. 'length'). Depending
on the component (image parameter) 'length' in $v$, the event $v$ proceeds to the left node 
(length $<$ split value) or to the 
right node (length $\geq$ split value) at the next lower level.
This node again splits in some other (or by chance the same) component, and the process continues. 
The result is that $v$ follows a track through the tree determined by the numerical values 
of its components, 
and the tree nodes' cut values, until it will end up in a terminal node.
This terminal node assigns a class label $l$ to $v$, which  can now be denoted 
as $l_i(v)$, where $i$ is the tree number.

\begin{figure}[h] 
\begin{center}
\includegraphics[totalheight=7cm]{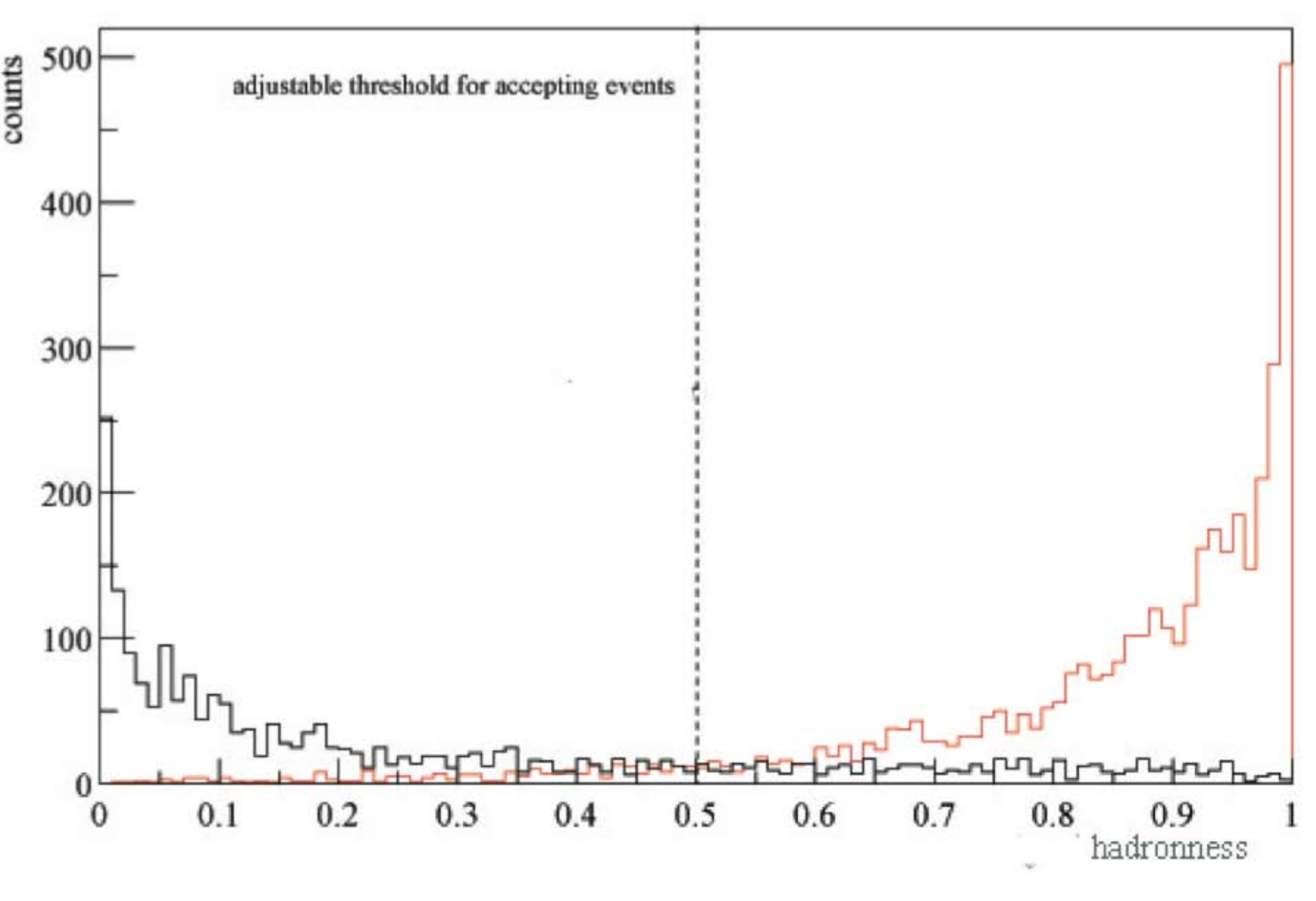}
\caption{{\it Mean hadronness for two test samples of gammas (left peak, black) and hadrons
(right peak, red). Hadronness is the final and only test statistic in $\gamma$/h separation.}}
\label{fig_had}
\end{center}
\end{figure}

The vector $v$ will be classified by all trees. Due to the randomization involved, 
different trees will often give different results, 
hence the name 'Random Forest'. From these results, a mean classification is calculated:
\begin{equation}
h(v) = \frac{\sum_{i=1}^{n_{trees}}l_i(v)}{n_{trees}}
\end{equation}
This mean classification is called Hadronness, and is used as the only test statistic (split-parameter) 
in the $\gamma$/h separation (see figure~\ref{fig_had}).

The splitting process is somewhat randomized by a feature called random split selection. The parameter 
candidates for a split are chosen randomly from the total number of available parameters. 
Among the candidates, the parameter and corresponding cut value to be used for splitting 
are chosen by the minimal Gini index. 
In the case of two classes, the Gini index $Q_{Gini}$ can be referred to as binomial 
variance of the sample 
scaled to the interval $[0, 1]$.
The Gini index (or GINI coefficient) can be expressed in terms of the node 
class populations $N_{\gamma}$, $N_h$ 
and the total node population $N$:
\begin{equation}
Q_{Gini} =  \frac{4}{N}\sigma_{binomial} =  4 \frac{N_{\gamma}}{N} \frac{N_h}{N} 
= 4 \frac{N_{\gamma}(N-N_{\gamma})}{N^2}   \in [0,1] 
\end{equation}
$Q_{Gini}$ of a node is zero for the ideal case that only one class is present in the node
($N_{\gamma}=0$ or $N_h=0$). The Gini index of the split is calculated by adding the
Gini indices of the two successor nodes (denoted by left and right node) and
scaling the result to [0,1]:
\begin{equation}
Q_{Gini} = 2  \left( \frac{N_{\gamma left}}{N_{left}} \frac{N_{h left}}{N_{left}} + 
\frac{N_{\gamma right}}{N_{right}} \frac{N_{h right}}{N_{right}} \right)    \in [0,1] 
\end{equation}

Choosing the smallest $Q_{Gini}$ corresponds to minimizing the variance of the 
population of gammas and hadrons, and naturally purifies the sample.
Minimization of the Gini index provides both the choice of the image parameter
and the split value to be used.

More details concerning the Random Forest method can be found in \cite{breiman2}. 
The original program 
was modified to calculate the mean hadronness instead of a $0$ or $1$ majority
vote for a class. Calculating the arithmetic mean by using 
weights (e.g. using the Gini index of terminal nodes) 
did not further improve the results \cite{bock},\cite{hengst}.

\section{Control of the training process}
\label{sec_control}
In this chapter we address some specific aspects of RF related to the training process. 
Proper training depends on several parameters, steering the growing of trees, 
which the user should be aware of. In the following these parameters are described.
\begin{itemize}
\item 
Number of trees:
the number of trees must be chosen large enough to ensure the convergence of the error $\sigma$, given by
\begin{equation}
\sigma(n_{tree}) = \sqrt{\frac{\sum_{i=1}^{n_{sample}}(h_i^{est}(n_{tree}) - h_i^{true})^2}{n_{sample}}}
\end{equation}
$\sigma(n_{tree})$ is the rms error of the estimated hadronness. $h_i^{est}(n_{tree})$ denotes the estimated 
hadronness (which depends 
on the number $n_{tree}$ of combined trees) and $h_i^{true}$ is the true hadronness
of event $i$ in the sample, which contains $n­_{sample}$ events in total.
The convergence process is shown in figure~\ref{fig_conv} for the training of 
RF on an MC gamma and MC hadron sample. 

\begin{figure}[h]
\begin{center}
\includegraphics[totalheight=6cm]{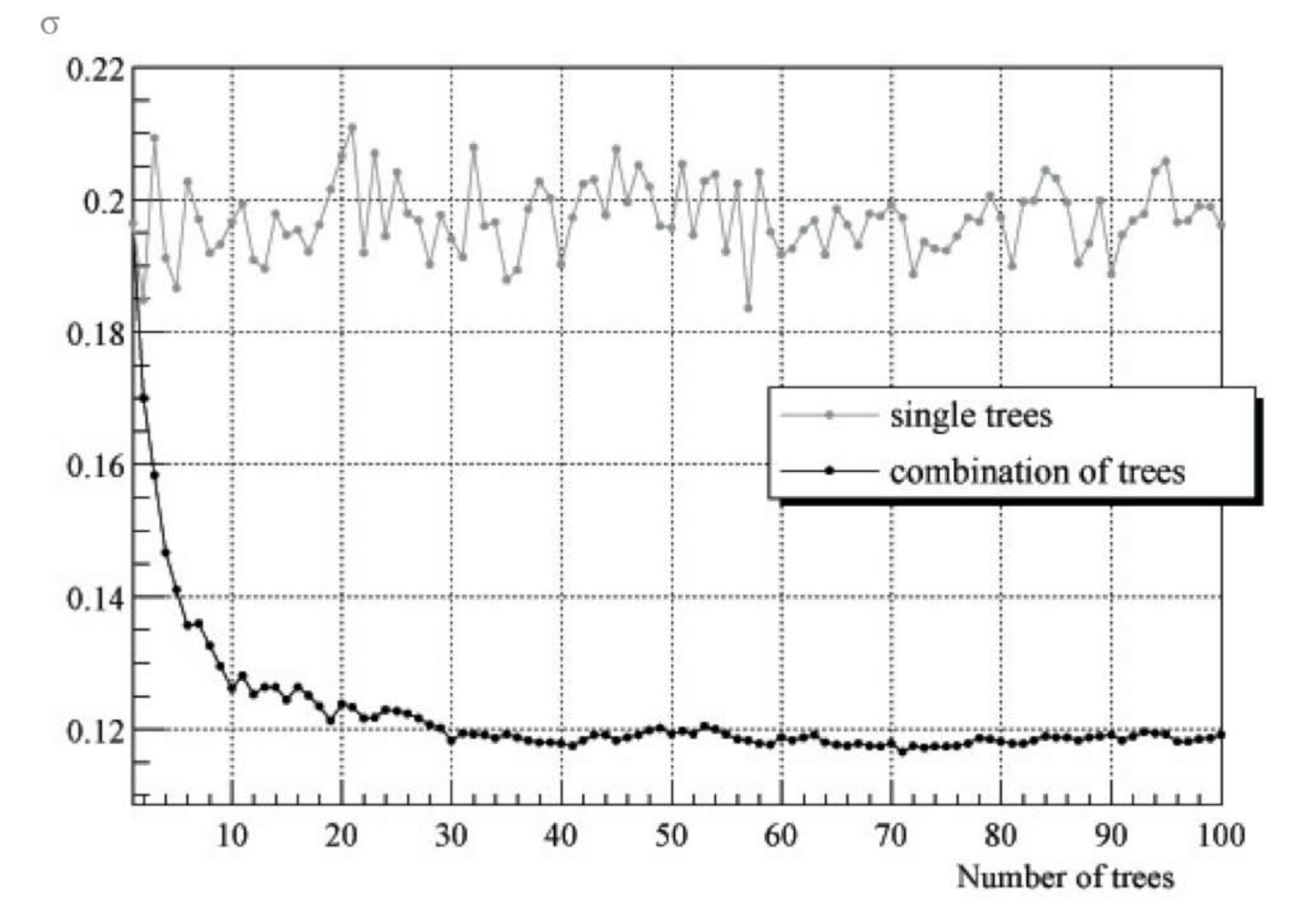}
\caption{{\it Error (rms, = $\sqrt(\sigma^2)$) of the estimated hadronness as function of the
number of trees used. Also shown is the variance of each single tree.}}
\label{fig_conv}
\end{center}
\end{figure}

Care was taken that the test sample, for which the figure was produced, 
is disjunct from the training sample. 
When taking events already used in the training process, $\sigma$ would be underestimated.
From figure~\ref{fig_conv}, the following practical method can be deduced:
One generates a reasonably high number of trees (100 trees is usually sufficient), performs the training process,
and then finds decisions for a test sample using a diminishing number of trees, to
judge how many trees still give satisfactory results. Trees generated during the training 
process are stored successively in a file. For the classification task one can read in the actually needed number of trees.
If no test sample is available, one can take $\sigma(n_{tree})$ as calculated from the so-called out-of-bag 
data during the training. 
The out-of-bag data are the 'residue' of the bagging procedure, as explained in the following. In the bagging procedure 
(generating of bootstrap samples, see chapter \ref{sec_basic})  there are data for each tree which have 
not been used for the tree's  bootstrap sample. Being independent, they can be used as test data for the corresponding tree. 
In other words, each event of the original training sample can be used as test data for $\approx 1/3$ of the trees.
If one observes a sufficient convergence of $\sigma$  calculated from out-of-bag data after, 
say, 150 trees, actually 50 trees are needed.
\item
Overtraining: During tree growing, the cut values of the parameters are adjusted according to the training sample. 
This overtraining is not a major drawback, it affects merely the training sample, which provides these 
exact cut values.
According to \cite{breiman2} the overtraining (or overoptimization) vanishes in case of an infinite number of trees. 
The practical method described above favours a minimal forest, with a number of trees sufficiently large to 
ensure a classification error (of a test sample), which is not significantly decreased by adding more trees.
Such a forest still shows overtraining: when applying $\gamma$/h separation to the training data, the classes of gammas 
and hadrons can usually be well separated by a cut in hadronness = 0.5. In other words, each tree 'learned by heart'
the training events, and the same is true for the entire forest. 
The situation is the same with classical cuts: the cut values are optimized on a certain 
observed data set from a gamma source or on Monte Carlo data, and later on applied to the data  
to be analyzed, which must not contain the training data.
\item
Number of trials in random split selection: This concerns the parameters considered
for splitting. A good empirical value for their number 
is $\sqrt{N}$ where $N$ is the total number of parameters used 
in tree growing \cite{breiman2}.
\item
Node size: this is the minimum size of node at which further splitting stops.
For correctly labeled training events $nodesize = 1$ can be used, for
partly incorrect labeled data (e.g. using ON-data as hadrons) $nodesize > 1$ is preferable, 
since data are not intended to be split completely. Experience tells that a small number $< 10$ is best.
\end{itemize}

\section{Application of RF in $\gamma$/h separation}
\subsection{Remarks concerning the training process}
In this chapter some features related to the Random Forest method will be briefly addressed. 
Some of these remarks are valid also for many other advanced classification methods in need of a training 
process, like Neural Networks or linear discriminant analysis.
\begin{itemize}
\item
Training data for Cherenkov telescopes:
We have used OFF data and MC gammas (correctly labeled samples) or ON data and MC gammas 
(partly wrongly labeled hadron sample). It is usually advisable not to use MC hadrons, 
since hadronic showers are 
difficult to simulate (unlike gamma showers which have a pure electromagnetic nature), 
so that MC hadrons are difficult to match in all details with 
real data. In fact, there is no need to use MC hadrons, when OFF or ON data are available.
Choosing ON data for training has the advantage of obviating OFF data taking, and of using data
taken under identical observational conditions. The
Random Forest algorithm is stable enough to deal with a hadron sample containing up to 1\% of gammas,
as shown in figure~\ref{fig_contam}, where the training was performed 
using OFF data with variable artificial contamination for the hadrons, 
and MC data for the gamma sample. 
\begin{figure}[h]
\begin{center}
\includegraphics[totalheight=7cm]{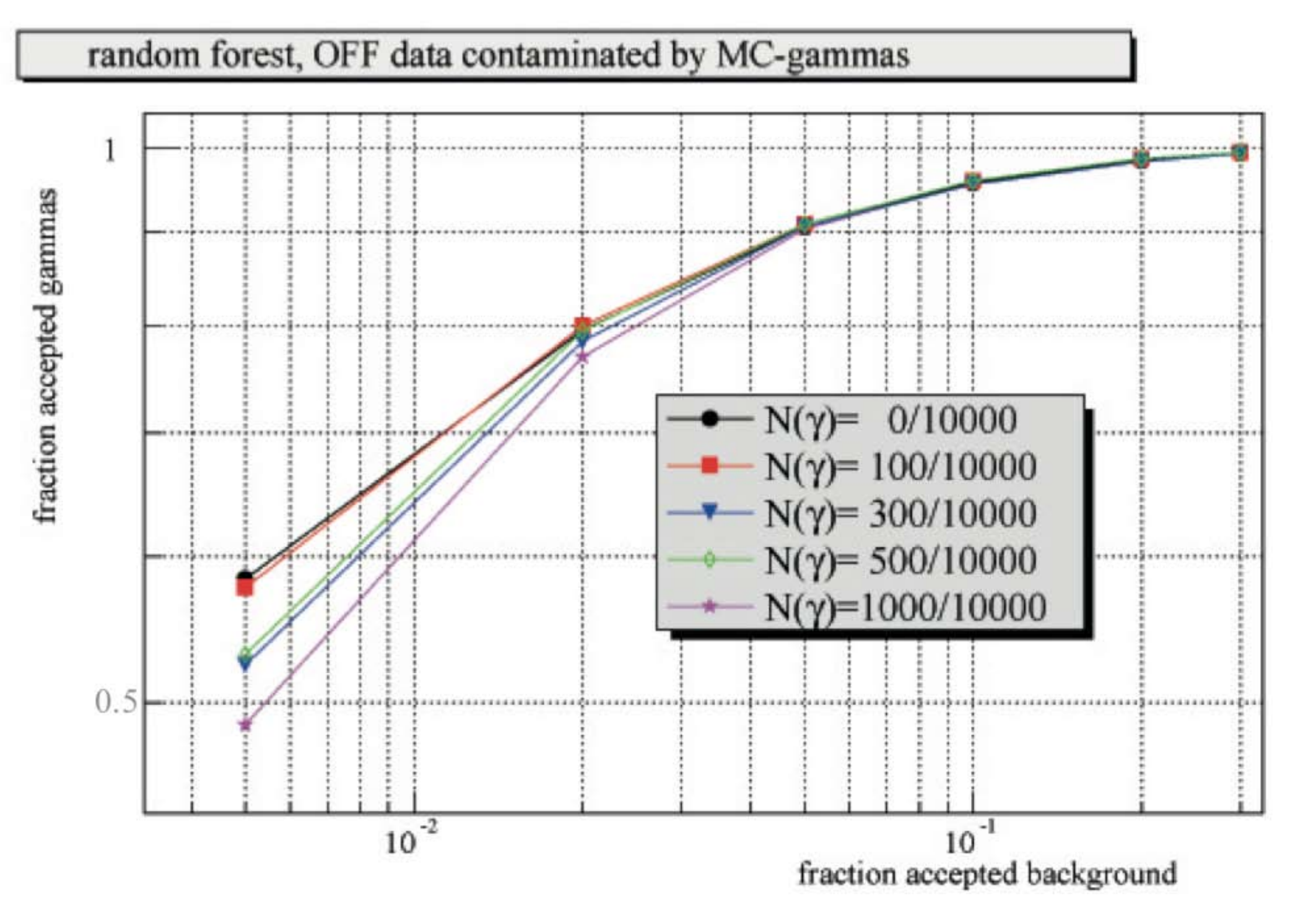}
\caption{{\it Neyman-Pearson or ROC diagrams of hadron training samples with 
a contamination of (mislabeled) gamma events. A hadron sample with 1\% gammas
introduces a negligible loss in selection efficiency.}}
\label{fig_contam}
\end{center}
\end{figure}
In order to simulate ON data, the OFF data were contaminated with MC gammas, i.e. the degree of 
contamination was known. For all simulated gamma admixtures the reduction of the
separation efficiency
beomes visible only in a region of low gamma acceptances, which is usually not advisible to 
operate in (too low gamma efficiency). Depending on the set of image parameters used for training,
a generalization of this result may not be possible. 
\item
Types of parameters:
All parameters are treated in the same way, which means that in particular detector-related or observational 
parameters like $cos(\theta)$ ($\theta$ being the zenith angle), $\bar{\sigma}$ (image noise, 
averaged over all pixels), or size  (integrated signal of the image),  must be 
used with care. The sense of using such parameters is that cuts in other image parameters will depend on them, 
but not that they should be used for cuts.
Thus, in  general, one can distinguish between parameters to be used for cuts, and 
parameters on which the cuts in other parameters may depend. 
To circumvent the problem, the training data must be chosen not to permit a classification using these parameters alone 
(e.g. by using the same (flat) distribution of $cos(\theta)$ in both training samples). 
Splits in these parameters, in training samples prepared this way,
can not directly serve for separating gammas and hadrons.
Additional attention must also be payed if e.g. the gamma data have discrete $cos(\theta)$ values for technical reasons 
in the Monte Carlo production. In this case the $cos(\theta)$ values appearing in the hadron sample must be 
rounded to the same values (binned), or the Monte Carlo data artificially spread to become continuous.
\end{itemize}

\subsection{Comparison with direct cuts in image parameters}
\label{results}
An extensive comparison of methods applied to Monte Carlo data sets for training and
test samples was given in \cite{bock}. One of the methods described there (called 
{\it Direct Selection}) was based on using simple AND/OR cuts in the multi-dimensional
space of image parameters. The choice of parameters or functions thereof
offers many possibilities for tuning.
We repeat here a similar comparison, again using Monte Carlo data, using {\it scaled} image 
parameters. Like in \cite{bock}, no claim can be made
that this result, found in favor of the RF method, 
can be generalized to all parameter choices or to real data.
Exhaustive comparisons with real data are lengthy, due to the high 
dimensionality of the problem,
which includes data selection and image cleaning steps even 
before image parameters are obtained.
Quality comparisons using real data are also influenced by the unavoidable changes in 
operation conditions, that are reflected in data corrections whose effect on separation 
methods are difficult to evaluate. A comparative study with comprehensive
MAGIC data samples is, however, in preparation.

\begin{figure}[h]
\begin{center}
\includegraphics[totalheight=7cm]{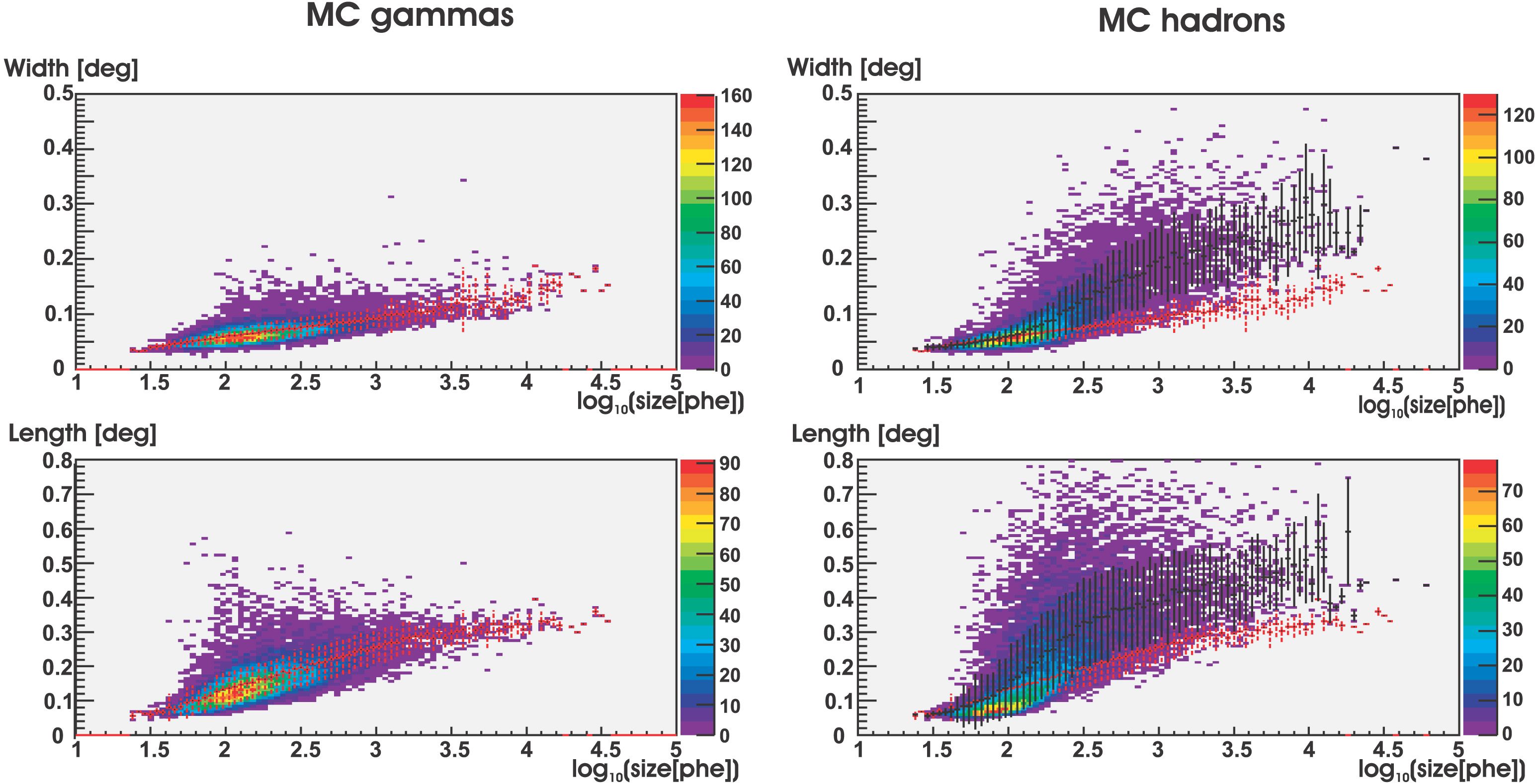}
\caption{{\it Distribution of the Hillas parameters width (top) and length (bottom) 
as function of log(size), 
for gammas (left) and hadrons (right), as used in the training samples. 
The profiles are shown in red (gammas) and black (hadrons), showing that
both parameters are good separators for size values above 200 photoelectrons 
(corresponding to about 100 GeV)}}
\label{fig_hill}
\end{center}
\end{figure}

For this comparison we used independent training and test samples, of 15000 events each. 
{\it Hadrons} were simulated with the parameters:
energy range $200GeV<E<30TeV$; spectral index $a = -2.7$; zenith angle range 
$0<\theta<30^\circ$; impact parameter range $0< R<400m$; viewing
cone $5^\circ$.
The {\it gamma} simulation settings were:
energy range $50GeV<E<30TeV$; spectral index Crab-like $a = -2.6$; 
zenith angle range $0<\theta<30^\circ$; impact parameter range $0<R< 200m$;
Figure~\ref{fig_hill} shows the corresponding 
distributions of the image parameters width [deg] and length[deg] as functions of
size [phe], for gammas and hadrons. 
All data were pre-cut to obtain high-quality training and 
test samples, requiring leakage\footnote{this parameter, not defined in \cite{bock},
uses an estimate of fractional energy escaping the camera} 
$<0.1$, dist$>0.3^\circ$, size $>200phe$.
Clearly, width and length are good separation parameters, at least for values of size 
exceeding 200 phe (photo electrons), 
which corresponds approximately to energies above $100GeV$. 
The size dependence of width and length can be dealt with
by using scaled parameters:
The size range (of MC gamma data) is divided into bins, and 
for each bin $i$ mean and variance of the 
width distribution ($\bar{w_i}$ and $\sigma^2_{w_i}$)
are calculated. The scaled width 
$w_{i,scaled}$ for each bin is then obtained by 
$w_{i,scaled} = (w_i - \bar{w_i}) / \sigma_{w_i}$.
 
The same procedure is used for the length parameter. As a result one obtains a normalized 
width and length distribution for gammas: they follow a pdf (probability density function) 
with mean 0 and variance 1.
In these variables, static (size-independent) cuts are used for $\gamma$/h separation. 
In order to find optimal cuts, a maximization of the $Q$-value which relates the relative 
acceptances of gamma-rays and hadrons
($Q = \epsilon_{\gamma} / \sqrt{\epsilon_h}$) 
was performed, using the Metropolis minimization package\footnote{which 
includes random perturbations in the search, thus avoiding to return local minima}
followed by a SIMPLEX minimization. Both packages are part of TMinuit
in the root analysis environment \cite{brun}. 

\begin{figure}[h]
\begin{center}
\includegraphics[totalheight=6cm]{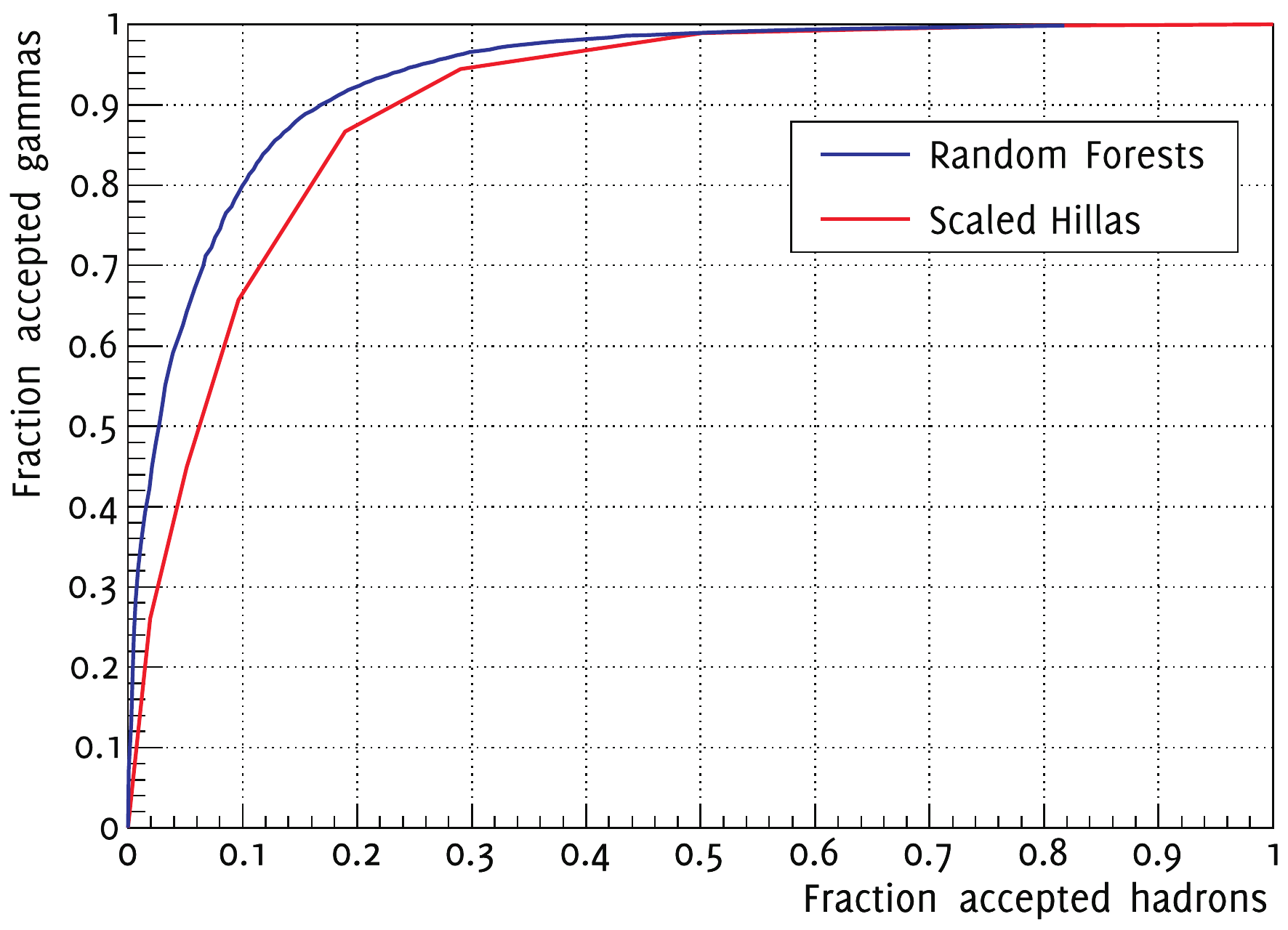}
\caption{{\it ROC curves for $\gamma$/h separation in the test sample, by the 
RF method (higher curve) and by cuts in scaled parameters, 
using the same parameters.}}
\label{fig_comp1}
\end{center}
\end{figure}

Both the Random Forest and the scaled parameter method used independent data
for training and testing. Only the parameters size, dist, width, and length were used.
The results are compared in the Neyman-Pearson or ROC (Receiver 
operator characteristic) diagrams of  figure~\ref{fig_comp1}; these diagrams
show gamma acceptance as function of hadron acceptance.
In order to obtain for the scaled parameter method more than a single point
(that of overall maximum $Q$) in the ROC diagram, a 
regularizer $a (\epsilon_h - p)^2$ was introduced (a generalization of 
the method used in \cite{bock}).
Here $p$ denotes a target acceptance for hadrons, and  $\epsilon_h$ is the freely variable 
hadron acceptance, which is obtained from the maximization of $Q$ and different for each $p$. 
We used a high scaling number $a = 1000$ to ensure that the optimization will give as a
result a set of cuts with  $\epsilon_h$ close to $p$. 

These results are shown as the lower curve in figure~\ref{fig_comp1}. 
We should stress again that this comparison can in no way show a general 
superiority of the RF method; practical experience shows that for a given 
data sample other methods (also including direct selection
as in the above example) can, at an effort, be fine-tuned to give results 
comparable to the RF method. However, in no case has the RF result been shown inferior,
and much less tuning is needed (and possible) with the RF method.
More comparisons (including also MAGIC data) can be found in \cite{zimmermann}.

\section{Using a Random Forest estimator for a continuous variable}
The RF method permits also to construct an algorithm of estimating a 
continuous quantity rather than a discrete class 
membership, dealt with in previous sections. We have used this method 
to estimate non-analytically the particle energy from the
measured image parameters. Two main approaches are possible:
\begin{itemize}
\item Forced division into classes:
Class labels are assigned to the training events
according to an energy grid. As a result, multiple classes
$E_0, E_1, ...,E_{n-1}$ are created. 
In the RF training process the related class populations are taken into account
together with a more general Gini index \cite{breiman1}
\begin{equation}
p_i = N_i / N
\end{equation}
\begin{equation}
Q_{Gini} = 1 - \sum_{i=0}^{n-1}p_i^2
\end{equation}
Here $i$ is the class index ($0 \leq i \leq n-1$). As already shown above, the Gini index of
a split is evaluated as sum of the two Gini indices obtained after the split, and minimized.
After the training procedure, the class populations
inside a terminal node are used to calculate the 
estimated energy  corresponding to the terminal node:
\begin{equation}
E_{est} = \frac{\sum_{i=0}^{n-1}E_iN_i} {\sum_{i=0}^{n-1}N_i}
\end{equation}
In this application of RF each tree returns an estimated energy and the overall mean
is calculated as the final estimated energy.

\item
A splitting rule based on the continuous quantity:
It is possible to completely avoid the use of classes by introducing a splitting rule,
which does not rely on class populations.
The idea of the Gini index (with its interpretation as binomial variance of the
classes) as split rule is a purification of the class populations, i.e. a separation
of the classes, in the subsamples after the split process. Similarly, when using the
variance in energy as a splitting criterion, the subsamples are purified with respect to
their energy distribution.
\begin{equation}
\sigma^2(E) = \frac {1}{n-1} \sum_{i=1}^{N}(E_i-\bar{E})^2 = 
\frac{1}{n-1} \left[ \left( \sum_{i=1}^{N}E_i^2\right) - n\bar{E^2}\right].
\end{equation}
In analogy to the Gini index of the split, the variance of the split is calculated by
adding the subsample energy variances, taking into account the node populations
as weights:
\begin{equation}
\sigma^2(E) = \frac{1}{N_L+N_R}\left(N_L\sigma_L^2(E) + N_R\sigma_R^2(E) \right)
\end{equation}
\end{itemize}

\begin{figure}[h]
\includegraphics[totalheight=4.5cm]{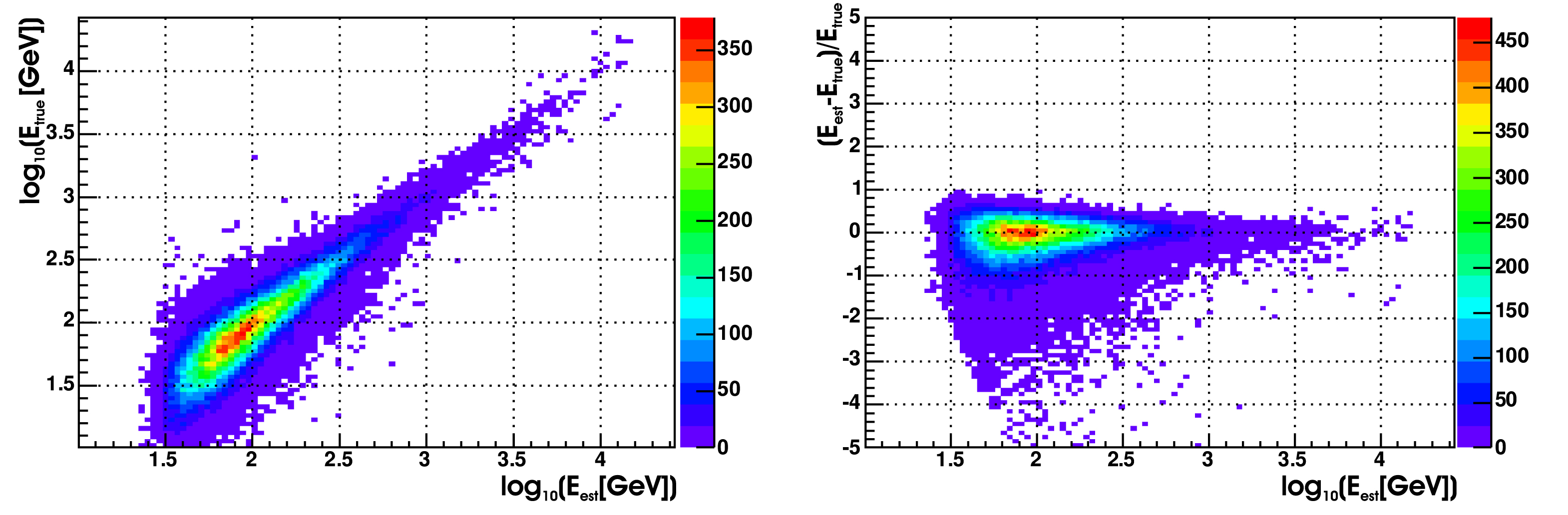}
\caption{{\it Left: The relation between the RF-estimated energy (horizontal)
and initial Monte Carlo energy (vertical axis) is perfectly linear.
Right: The rms error $\sigma_E/E$ as function of initial energy.}}
\label{fig_energy}
\end{figure}

We have used both approaches for a set of Monte Carlo data. 
With 100 classes for the first (classification) method,
it produces results nearly identical to those of the second (regression) approach.
The results of this latter RF approximation for energy
can be seen from figure~\ref{fig_energy}.
The linearity is perfect, and the energy resolution (as defined by
the rms error $\sigma_E/E$) comes out 26\% at 100~GeV and
19\% at 1~TeV,
very fair values for a single telescope (telescope arrays can reach better resolution).
We have not found an analytical parameterization for energy expressed in terms of image 
parameters giving a result better than with the RF representation; with extensive tuning,
results comparable in quality have been found, though.

\section{Conclusions}
The Random Forest (RF) method based on multiple decision trees 
was extensively tested as an analysis tool in the $\gamma$/h separation 
for data obtained with the MAGIC telescope.
In this paper we discuss many implementation details and the 
parameters a user has to become familiar with.
We also compare the performance of RF with the more
conventional technique of cuts in scaled image parameters, using MC
data. It could be shown that RF in this comparison is superior
to the classical method. This comparison does not
imply a general superiority of the RF method; practical experience
shows that for a given data sample the conventional methods (like
dynamical cuts or cuts in scaled image parameters) may be tuned
to give results comparable (but not superior) to the RF method.
A dedicated comparative
study using MAGIC experimental data is still under way. 

The RF method does produce stable results and
is robust with respect to input parameters, even if strongly correlated. The method
adjusts itself to the available multi-dimensional space, 
with a minimum of human intervention:
there are only few tunable parameters, which can be chosen according to simple criteria 
(number of trees, trials in random split selection and final node size). 
This simpler control and tuning can then be seen as a general advantage
over conventional methods.
Proper training samples, however, are important, as in any advanced
method requiring a training process, i.e. 
one has to rely on a good Monte Carlo simulation. Using OFF or ON data as hadron 
sample limits the MC dependence to the gamma showers, better understood 
than hadron showers. 
There remains, however, the need to correctly treat 
atmospheric conditions under different zenith angles, 
and good knowledge of the detector. 

Training and classification are fast: benchmarks using
a 1.5~GHz PC (Athlon XP), with training and test samples each containing 10.000~events, 
a total of 10~image parameters used,
100~trees used for classification, each tree completely grown (nodesize=1), 
3~trials in random split selection, give one minute for training and 2~ms/event for classification.
A comparable analysis technique like Neural Networks demands substantially 
more computer time for training.


\section*{Acknowledgement}
We thank Jens Zimmermann for fruitful discussions about the RF method
and for  comparisons of the RF method with a Neural Net approach.




\begin{thebibliography}{10}
\bibitem{hillas}A.M.Hillas: Proceedings of the 19th International 
Cosmic Ray Conference, ICRC 1985 La Jolla , 3 (1985) 445
\bibitem{hillas1}A.M.Hillas: Space Science Rev. 75 (1996) 17
\bibitem{fegan}D.J.Fegan: J.Phys.G, Nucl.Part.Phys. 23 (1997) 1013
\bibitem{aharonian}F.Aharonian et al.: Astropart.Phys. 6 (1997) 343
\bibitem{kraw}H.Krawczynski et al.: Astropart.Phys. 25 (2006) 380
\bibitem{bock}R.K.Bock, A.Chilingarian, M.Gaug, et al., 
Nucl. Inst. and Methods A 516 (2004) 511
\bibitem{hengst}T.~Hengstebeck, PhD thesis,
Mathematisch-Naturwissenschaftliche Fakult\"at I,
Humboldt-Universit\"at zu Berlin, M\"arz 2007.
Available at URL http://edoc.hu-berlin.de/docviews/abstract.php?id=28015
\bibitem{lorenz} E. Lorenz, New Astron. Rev. 48 (2004) 339
\bibitem{breiman1}L . Breimann, J. H. Friedmann, R. A. Olshen, C. J .Stone: 
Classification and Regression Trees, Wadsworth, 1983
\bibitem{albert1} J.Albert et al., Astroph. Journal 664 (2007) L87
\bibitem{albert2} J.Albert et al., Astroph. Journal 665 (2007) L51
\bibitem{albert3} J.Albert et al., Astroph. Journal 669 (2007) 1143
\bibitem{albert4} J.Albert et al., to be published in Astroph. Journal,
preprint available at http://de.arxiv.org/abs/0705.3244
\bibitem{breiman2}L.Breiman, FORTRAN program Random Forests, Version 3.1, and
L.Breiman, Manual On Setting Up, Using, And Understanding Random Forests V3. 1, 
both available at http://oz.berkeley.edu/users/breiman
\bibitem{brun}R.~Brun, F.~Rademakers, http://root.cern.ch/
\bibitem{zimmermann} J.~Zimmermann, PhD thesis, Fakult\"at f\"ur Physik,
Ludwig-Maximilians-Universit\"at M\"unchen, Juni 2005.
Available at URL http://edoc.mpg.de/274832


\end{thebibliography}
\end{document}